\documentclass[icps3]{svjour}

\usepackage{latexsym}
\usepackage{graphics}
\usepackage{amsmath}   
\usepackage{epsfig,psfrag}

\begin{document}
\title{Phonons and Phonon Confinement in Transport through Double
       Quantum Dots}
%
\titlerunning{Phonons and Phonon Confinements in Double Quantum Dots}
%
\author{S. Debald\inst{1} \and T. Vorrath\inst{1} \and T. Brandes\inst{1,2}
\and B. Kramer\inst{1}
}                     
%
\authorrunning{S. Debald et al.}
%
\institute{I. Institut f\"ur Theoretische Physik, Universit\"at Hamburg,
           Jungiusstr. 9, 20355 Hamburg, Germany,
           \email{debald@physnet.uni-hamburg.de}
           \and Department of Physics, University of Manchester Institute of Science and Technology (UMIST), P.O. Box 88, 
Manchester M60 1QD, United Kingdom
           }
\maketitle
\begin{abstract}
We calculate the electron--phonon interaction coefficients for surface
acoustic waves and for phonons in free standing quantum wells. These
are used to derive the inelastic current through a double quantum dot
caused by spontaneous emission of phonons. For the case of the free
standing structure (phonon cavity), we predict a staircase--like
inelastic current superimposed by van Hove singularities. Therefore,
the phonon confinement can be detected by electron transport
measurements.  
\end{abstract}
\section{Introduction}
\label{intro}
The fabrication of free--standing or suspended nanoscale structures
\cite{Rou99} has opened new perspectives for the investigation of
mesoscopic phonon effects such as phonon confinement, thermal
conductance quantization, or single phonon transport.  Recent
experiments are designed to show and emphasize mesoscopic phonon
effects of few or even a single phonon \cite{SchEtal00}.  At the same
time, sensitive detectors for phonons become important
\cite{AguKou00}. Electron transport measurements have already been
used successfully for this task.

In double quantum dots, the emission of phonons has been observed to
dominate the non--linear electron transport at mK--temperatures
\cite{FujetalTaretal}.  This effect could be explained by an X--ray
singularity--like boson shake--up effect for tunneling of single
electrons due to coupling to phonons \cite{Brandesddot}.  Different
phonon modes (surface, bulk, or confined) and features in phonon
densities of states can therefore (at least in principle) be
identified in electronic transport measurements.  The energy
difference between the two dot ground states should be used to scan
the relevant phonon energy window.
\section{Surface Acoustic Phonons}
\label{saw}
The knowledge of the electron phonon interaction coefficient is
necessary in order to quantitatively understand the influence of
different kinds of phonons in electron transport measurements. Here,
we consider surface acoustic waves (SAW) and their interaction with
electrons.  In most experiments, the relevant electrons are part of a
2DEG which is located a small distance $d$ beneath the surface of the
sample (typically $d\approx 100$nm).  If the SAW wavelength becomes
comparably small, it is important to take into account the
non--monotonic decrease of its amplitude into the sample.  For an
anisotropic crystal like GaAs, we calculate the piezoelectric
electron--phonon interaction coefficient as a function of the depth
and compare it with results for an isotropic crystal. Finally, we
apply the result to estimate the inelastic tunnel current through a
double quantum dot, caused by emission of SAW.
\begin{figure} 
\centering
\psfrag{energie}{\small $\omega / \omega_d$}
\psfrag{strom}{\small $\rho(\omega)\omega_d$}
\psfrag{qz}{\small $qz$}
\psfrag{gamma}{\small \hspace*{-6mm}$\gamma_q /10^{-13}$ eVm}
\epsfig{file=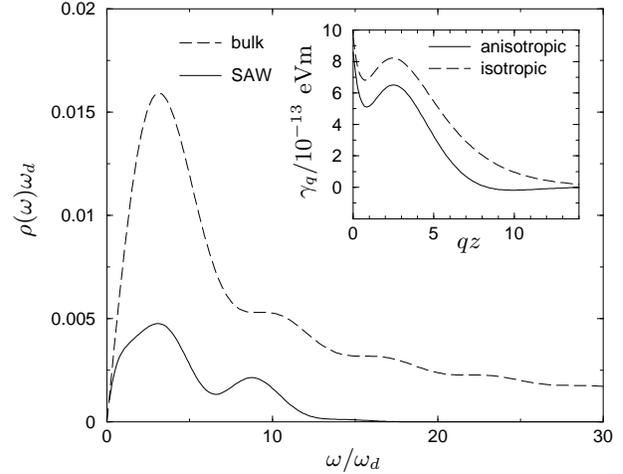, width=8cm}
\caption{Effective density of states $ \rho(\omega)$ as a function of
the energy $\omega$ in units of $\omega_d = v/d$, with $v$ the SAW
velocity and $d$ the distance between the two quantum dots. We chose a
value of $d=250$nm and a distance of 100nm from the dots to the
surface.  The inset shows the electron--phonon interaction coefficient
$\gamma_q$ as a function of $qz$ with wavenumber $q$ and depth $z$ in
the anisotropic and the isotropic model of the crystal.}
\label{rho}
\end{figure}

We consider a SAW propagating in [110] direction on the [001] surface
of an anisotropic cubic crystal. This wave causes an electric
potential $\varphi$ by the piezo effect. Hence, electrons in a 2DEG or
a quantum dot close to the surface interact with the SAW as they are
exposed to the potential $V=e \varphi$. The interaction potential can
be written in the common form
\begin{equation}
V_{\rm int} (x,z)= \frac{1}{L} \sum_q \, \gamma_q(z) \, e^{iqx} \, (b_q +
b_{-q}^\dag)
\end{equation}
with wave number $q$, normalization length $L$ and boson annihilation
and creation operators $b_q$, $b_q^\dag$. $\gamma_q$ is the
interaction coefficient, that is $z$--dependent for the case of
surface waves.  Calculating the piezoelectric potential \cite{simon},
we find for $\gamma_q$ the expression
\begin{equation}
\label{int-coef}
\gamma_q (z) = \,\tilde{C}\;\{A_1 e^{ib_1qz} + A_2 e^{ib_2qz} + A_3 e^{-qz} \} 
\end{equation}
where all prefactors are combined in $\tilde{C}$ which does not depend
on $q$. $A_1, A_2, A_3, b_1, b_2$ are complex functions of the
material constants.  For GaAs we find for $\gamma_q$ on the surface a
value of $8.6 \cdot 10^{-13}$ eVm. This is slightly smaller as the
value found in the isotropic model of the crystal
\cite{knaebchen}. The dependence on the depth is shown in the inset of
Fig. \ref{rho}.  It can be seen that the isotropic model slightly
exaggerates the penetration depth of the interaction coefficient.

We use the interaction coefficient (\ref{int-coef}) to estimate the
inelastic tunnel current through a double quantum dot caused by
spontaneous emission of surface acoustic pho\-nons. Such a current was
measured in recent experiments \cite{FujetalTaretal}.  Following the
theoretical approach in \cite{Brandesddot}, we calculate an effective
density of states $\rho(\varepsilon)$ which serves as an approximation
for the inelastic tunnel current.  The result is shown in
Fig. \ref{rho} as well as the result for interaction with bulk phonons
from \cite{Brandesddot}.  Both results possess the oscillations in the
inelastic current that were found in the experiment, although the
inelastic current in the case of emission of bulk phonons is stronger.

\section{Confinement in Phonon Cavities}
In (partly) freestanding structures \cite{Blickneuespaper}, electron
transport through lateral double dots is expected to show signatures
of phonon confinement.  We investigate a free standing quantum well of
finite thickness as a phonon cavity model. Due to the confinement, a
quantization of the phonon modes similar to the electronic case
occurs. Families of shear, flexural and dilatational modes can be
classified as in classical acoustics.

\begin{figure}[h]
\begin{center}
\psfrag{w}{\small $\omega / \omega_b$}
\psfrag{i}{\small $I_{\rm inel}(\omega)$}
\psfrag{n}{\small $\nu(\omega)$}
\psfrag{a}{\small $\nu(\omega)$}
\psfrag{b}{\small $I_{\rm inel}(\omega)$}
\epsfig{file=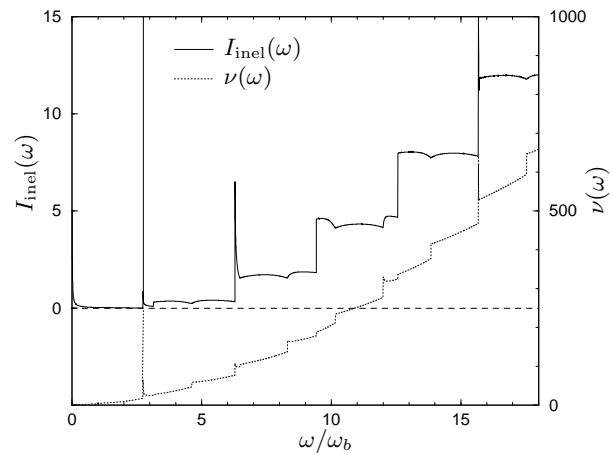, width=8cm}
\caption{Thermodynamical density of states of the cavity phonons
$\nu(\omega)$ (dashed) and inelastic current through the double dot in
the cavity (solid). The 2$^{\rm nd}$ and 13$^{\rm th}$ flexural modes
lead to van Hove singularities in the density of states and the
inelastic current. For GaAs and a cavity width of $2b = 1\mu$m the
characteristic energy is $\hbar \omega_b =\hbar c_l/b= 7.5 \mu e$V. }
\label{fig:cavity}
\end{center}
\end{figure}

We calculate the dispersion relation of the cavity phonons using a
numerical approach and find a staircase-like phonon density of states
$\nu(\omega)$ superimposed by van Hove singularities as shown in
Fig. \ref{fig:cavity}. The characteristic energy scale is given by
$\omega_b = c_l / b$ with the longitudinal velocity of sound and the
cavity width $2b$.

For small enough cavity confinement widths one can show that the
deformation potential is the dominant interaction mechanism between
electrons and confined phonons. For a cavity confined in $x$ direction
the deformation potential is \cite{BanEtal}
\begin{equation} \nonumber
V_{\rm def} = \sum_{{\bf q}_{\|},n} \gamma_n({\bf q}_{\|}) e^{i {\bf
q}_{\|} {\bf r}_{\|}} \left\{ \begin{matrix} \cos{q_{l,n} x} \\ \sin{
q_{l,n}x} \end{matrix}\right\} \left[ b_n({\bf q}_{\|}) +
b_n^{\dagger}(-{\bf q}_{\|}) \right],
\end{equation}
where ${\bf q}_{\|}$ and $q_l$ are the in--plane and transversal
components of the wavevector, resp. The upper row is for dilatational
and the lower is for flexural modes. Dilatational phonons induce a
symmetrical potential with respect to the cavity midplane, flexural
phonons an antisymmetrical potential. Shear waves do not induce a
deformation potential.
%
%
For a double quantum dot placed in the middle of the cavity we
calculate the inelastic current in the same model as in Sec. 2 using
the deformation potential interaction between electrons and confined
phonons. We find that due to the symmetry of the different phonon
families, one can suppress the emission of such phonons.  This can be
done by changing the double dot axis relative to the confinement
direction of the phonon cavity. Therefore, one can modify the emission
characteristics of the double dot.  Furthermore, we calculate the
inelastic current through the double dot caused by flexural phonons
corresponding to the emission characteristic of the dots oriented in
direction of the confinement. We find that the traces of phonon
confinement seen in the thermodynamical density of states remain in
the inelastic current as shown in Fig. \ref{fig:cavity}. For a phonon
cavity made of GaAs and a width of $2b = 1 \mu$m, the characteristic
energy is $ \hbar \omega_b = 7.5 \mu e$V. Therefore, it should be
possible to detect the phonon confinement in electron transport
experiments through coupled quantum dots.

\end{document}